\documentclass[reprint,amsmath,amssymb,aps]{revtex4-1}

\usepackage{graphicx}%
\usepackage{dcolumn}%
\usepackage{bm}%
\usepackage{hyperref}
\usepackage{cleveref}
\usepackage{xcolor}

\newcommand{\Isotope}[2]{\ensuremath{{}^{#2}\mathrm{#1}}}

\newcommand{\neutrino}[1]{\ensuremath{\nu_{#1}}}
\newcommand{\antineutrino}[1]{\ensuremath{\overline{\nu}_{#1}}}

\def\cevns{\ensuremath{\mathrm{CE\nu NS}}}
\def\msolar{\ensuremath{\mathrm{M_{\odot}}}}

\def\keV{\ensuremath{\mathrm{keV}}}
\def\MeV{\ensuremath{\mathrm{MeV}}}

\def\mus{\ensuremath{\mathrm{\mu s}}}

\begin{document}

\title{Supernova Detection at SNOLAB}
\thanks{Thanks to CJP for proposing this special collection.}%

\author{Erica Caden}
\email{erica.caden@snolab.ca} 
\affiliation{SNOLAB}%

\author{Stephen Sekula}
  \altaffiliation[Also at ]{Queen's University}
\affiliation{SNOLAB}

\author{Stanley Yen}
\affiliation{TRIUMF}%

\date{\today}%

\begin{abstract}
Neutrinos carry most of the energy released by a core-collapse supernova.  SNOLAB has two neutrino-capable detectors, SNO+ and HALO, that have complementary neutrino flavour sensitivities. SNOLAB is also host to existing facilities, or plans to host future projects, that can enhance sensitivity to these neutrinos. These detectors, together with others worldwide both in existence and planned, will provide insights to a variety of different models using neutrinos from the next galactic supernova.

\end{abstract}
\maketitle

\section{\label{sec:intro}Introduction to Supernovae }

A core-collapse supernova (CCSN) occurs in a galaxy like the Milky Way, on average, once or twice per century ~\cite{Rozwadowska:2020nab,Tammann:1994ev}. 
These explosions are a spectacular sight in visible light, where, for a few days, the supernova rivals the entire host galaxy's $\approx$100 billion stars in brightness. 
A CCSN is a cosmic machine for turning gravitational potential energy primarily into neutrinos. 
Most (99\%) of the energy released by a CCSN is carried away by neutrinos, 1\% by the kinetic energy of the shock wave, and only 0.01\% as optical radiation.  
Visible light observed on earth comes only from the outer fringes of the explosion, where the matter is dilute enough for photons to propagate freely toward the Earth. 
Thus the visible light emerges late in the explosion when the outgoing shock wave has propagated to the surface.  
Stellar matter is, by contrast, transparent to neutrinos, and neutrinos therefore give us a prompt signal of the nuclear processes occurring in the interior of the star during the core-collapse and explosion.  
The observation and study of neutrinos from supernovas is essential to a complete understanding of the stellar life cycle.

The stages of stellar evolution leading to a CCSN are as follows.  A CCSN is the cataclysmic endpoint in the life of a star whose initial mass exceeds eight solar masses ($8~\msolar$). The main sequence stage of stellar evolution, during which the star is powered by hydrogen fusion in the core,  represents the longest stage in the life cycle of a typical star. This stage is maintained by an equilibrium between inward gravitational collapse and the outward pressure of radiant energy generated in the core by nuclear fusion. A main-sequence star slowly evolves away from its initial 75\% H, 25\% He composition as fusion reaction products accumulate in the core. As the hydrogen in the core is depleted, stars will move off the main sequence into cycles of collapse, new fusion burning, and expansion as the increasing temperature in the stellar core enables the fusion of increasingly heavier elements.  This eventually results in an onion-like structure, with an innermost core of Fe/Ni, surrounded by concentric shells of progressively lower-mass elements such as S, Si, Ne, O, C, and He at larger radii, with ordinary H fusion occurring only in the outermost layer.   The core accumulates iron-region nuclei near the peak of the binding energy per nucleon, the exact composition of which is determined by a competition between charged-particle capture and photodistintegration~\cite{Rolfs1988, Fewell1995}. There is now no avenue for further nuclear energy production in the core; the outward radiation pressure ceases, and gravitational collapse follows.

The pressure of the core is now due solely to the presence of a degenerate electron gas. 
If collapse is stopped by this pressure, then the star achieves the condition to become a white dwarf. 
The Sun (and stars like it) is expected to end in this stage.  However, in heavier stars where the mass of the Fe core exceeds the Chandrasekhar limit of 1.4~\msolar, electron degeneracy pressure is unable to resist the inward pull of gravity and core collapse is initiated. 
The collapsing core achieves an overall lower potential energy when protons (either bound in the Fe nuclei or liberated by photodisintegration) undergo  \emph{electron capture} (EC),
\begin{equation}
    e^- + p^+ \to n + \neutrino{e}. \label{eqn:EC}
\end{equation}
This reaction depletes electrons but generates neutrinos to conserve lepton number and results in the ``deleptonization'' $\neutrino{e}$ burst. 
This pulse accounts for $\sim 1\%$ of the total neutrino flux generated by a CCSN and is confined to a sharp luminosity spike of a few milliseconds.  

The depletion of electrons enables the core to collapse to nuclear matter density, forming a proto-neutron star with a diameter ~25,000 times smaller than the initial core. 
Material in-falling to fill the void powers the accretion stage, which is dominated by \neutrino{e} and \antineutrino{e} emission. 
Even though EC is endothermic, the energy loss is more than compensated by the much lower gravitational potential energy of the compacted core. The released gravitational energy heats the core, raising the temperature to the point where neutral current (NC) scattering processes such as
\begin{equation}
e^+ + e^- \rightarrow Z^{0*}  \rightarrow \neutrino{} + \antineutrino{}
\end{equation}
prodigiously produce neutrino-antineutrino pairs of all flavours equally.  Since neutrinos interact more weakly with the surrounding stellar matter than other particles, it is neutrino emission that is most effective in cooling off the proto-neutron star. This cooling phase accounts for the bulk of the neutrino emission in a core-collapse supernova. 

The density of matter in the hot proto-neutron star is such that the absorption length of a $\sim 10\MeV$ neutrino is $\leq 100~\mathrm{m}$ and decreases with increasing energy~\cite{Reddy:1997yr,Shen:2003ih,Rrapaj:2014yba}. The neutrinos are thus trapped in the proto-neutron star and thermalize. %
Since electron neutrinos interact with matter more strongly than electron antineutrinos (which in turn interact more strongly than neutrinos of other flavours) the electron neutrinos are radiated from a surface of last scattering---the $\neutrino{e}$ \emph{neutrinosphere}---that is further from the centre of the proto-neutron star and hence at a lower temperature.  Thus, a hierarchy of neutrino temperatures is a generic and robust prediction of supernova models:
\begin{equation}
T(\neutrino{e}) < T(\antineutrino{e} ) < T( \neutrino{x} ).
\end{equation}

Detailed calculations of the time evolution of the neutrino luminosity and energy for each of the different flavours as a function of time have been performed~\cite{Nagakura22}.  The ability to view each neutrino flavour's energy as a function of time would facilitate the measurement of core temperature at different depths and verify the fundamental processes of core collapse and cooling. The shape of the neutrino spectrum is not precisely Fermi-Dirac (FD), due to the energy-dependent opacity of neutrinos in the outer regions of the proto-neutron star. There are microscopic processes involved in neutrino transport~\cite{Keil:2002in,Raffelt:2001kv}. 
These result in a ``pinched'' spectral shape where the high-energy tail is suppressed relative to a pure thermal spectrum of the same mean energy. 
The observation of the spectral shapes, for the different neutrino flavours, is a fundamental test of our understanding of neutrino opacity at nuclear-matter densities.   Reviews of these subjects are available in Refs.~\cite{Janka:2012wk,Raffelt:2007nv,Janka:2017vlw,Suzuki:2024fse}. 
The $\neutrino{e}$ and $\antineutrino{e}$ spectral shapes will be measurable with the current world-wide suite of detectors. 
Most interesting would be the observation of an abrupt cutoff in the neutrino flux induced by the formation of a black hole;  the shape of the cutoff could contain information on the mass of the proto-neutron star ~\cite{Wang:2021elf} or neutrino``echo'' off the infalling matter ~\cite{Gullin:2021hfv}.

Most of the dynamics of a CCSN occur between the collapse of the core and the explosion of the star. Stellar cores that just exceed the Chandresekhar limit are no longer supported by electron degeneracy pressure. Stellar matter then accelerates towards the proto-neutron star and reaches velocities approaching $0.25c$. 
The collapsed core is described by a rigid equation of state. 
When the infalling matter encounters this core it is redirected back outward.
In most realistic 2-D or 3-D models the outward-moving shock wave eventually stalls, its momentum sapped by photodisintegration of nuclei and collision with the infalling material. 
It is widely believed that the shock wave is revived by neutrinos interactions in the densest area of the shock density profile, but there exist some models with alternate explanations \cite{Vartanyan:2023zlb}.  
The situation is complicated by turbulence, asymmetries, the possible presence of standing accretion shock instabilities, as well as the details of microphysics and neutrino transport, matter-induced and neutrino-neutrino-induced neutrino flavour transformations, the actual nuclear equation of state, endothermic processes involved in nucleosynthesis, etc. Some of these processes are expected to imprint distinct signatures on the neutrino signal. This reinforces the importance of observing CCSN neutrinos to better understand these effects.  

Once successfully revived the shock wave may take (depending on the size of the progenitor) from 30 minutes to 10 hours to reach the stellar surface and violently eject the outer layers of the star.  This is when the optical radiation is produced, accounting for only $\sim ~0.01\%$ of the total energy liberated by the supernova.  Neutrinos, by virtue of their promptness, not only give a real-time signal of the nuclear and particle processes of the core-collapse, but also provide an ``early warning'' that can be used to alert the astronomical community to an impending electromagnetic signal and facilitate very early observations of the supernova.  This is the objective of the Supernova Early Warning System, SNEWS (see \Cref{sec:snews}), a global network of co-operating neutrino detectors that correlate possible supernova neutrino signals between detectors, allowing an automated alert to be sent to astronomers with high confidence~\cite{Antonioli:2004zb,AlKarusi:2021}.  

\section{\label{sec:level1} Neutrino Detectors and Supernovae }

Data from the detectors at SNOLAB (\Cref{sec:snolab}) and other facilities would permit, for the first time, a flavour decomposition of the neutrino flux (identification of the separate contributions of $\neutrino{e}$ , $\antineutrino{e}$ , and $\neutrino{x}$,  where $x=\mu\textrm{ or }\tau$) from a galactic supernova.  This is possible because  the detection in \emph{hydrogenous detectors} (those based on scintillator and/or water targets) is primarily $\antineutrino{e}$-sensitive. In contrast, the detection in argon (Ar) and lead (Pb) detectors is primarily $\neutrino{e}$-sensitive. The detection of neutral current processes, like the excitation of the 15.11 MeV state in \Isotope{C}{12}, is equally sensitive to all flavours of neutrinos and antineutrinos. A flavour decomposition is important because neutrinos can undergo flavour oscillations between the point of production and the detector. While this complicates the interpretation of detection it also provides an opportunity to observe otherwise inaccessible neutrino phenomena. 

The neutrino oscillations are of three types: vacuum, matter-induced, and neutrino-induced.  Vacuum oscillations occur even in space free of external matter. Neutrinos are generated by weak processes and thus begin as flavour eigenstates, denoted $\neutrino{e}$, $\nu_{\mu}$, or $\nu_{\tau}$. The weak eigenstates are not required to be the mass eigenstates, in which case each set of basis states can be expressed as superpositions of the other (linear combinations of the mass eigenstates $\nu_1$, $\nu_2$, and  $\nu_3$).  The observed oscillations of solar, reactor and atmospheric neutrinos imply that at least two of the mass eigenstates must have non-zero masses. 

Matter-induced oscillations (also known as the Mikheyev–Smirnov–Wolfenstein, or MSW, effect) occur because the electrons in matter exert a different weak potential on $\neutrino{e}$ than on other flavours. At typical supernova neutrino energies of $\sim$10 MeV, charged current (CC) interactions like $\neutrino{x} + e^- \to \ell_x + \neutrino{e}$ involving atomic electrons are energetically forbidden for the non- $\neutrino{e}$ flavours.  For particular values of energy and electron density, this potential induces a resonant transition from one flavour to another.  It is anticipated that there are two such resonances in the density profile of a supernova.  

In addition, the MSW effect inside the star turns the neutrino into a mass eigenstate at the resonant local density of matter. Assuming an adiabatic transition to zero matter density, the neutrino emerges from the stellar material as a pure mass eigenstate in vacuum. Such a state would then undergo no further flavour oscillations between the supernova and the earth~\cite{dosSantos:2023skk}. However, once such a state arrives at Earth there can again be matter-enhanced flavour transitions that will modify flavour composition of the burst at different geographic locations. This further necessitates the comparison of measurements in independent detectors in order to gain information on oscillation parameters~\cite{Borriello:2012zc}.

Neutrino-induced oscillations occur when  the density of neutrinos trapped in the core of a supernova is so high that the neutrinos scatter off each other. Thus, each neutrino feels a potential due to the other neutrinos.  This induces MSW-like flavour oscillations, but due to $\nu\text{--}\nu$ rather than $\nu\text{--}e$ interactions~\cite{Duan:2010bg}. Nowhere else in the universe is there a sufficient density of neutrinos to observe such collective $\nu\text{--}\nu$ interactions, which cause the whole ensemble of neutrinos to collectively undergo synchronized oscillations in flavour space. This behaviour is analogous to the precession and nutations of a pendulum. The observable effects of these collective oscillations is still a topic of active research ~\cite{Sasaki:2019jny}.

The neutrinos that arrive at Earth will interact with matter in a state of definite flavor, either in $\neutrino{e}$, $\antineutrino{e}$, $\neutrino{x}$, or $\antineutrino{x}$. The probability of seeing any of these in the initial state of a reaction will depend on the mixing history of the original neutrinos. A few reaction channels are available to the neutrinos. These include the \emph{inverse beta decay} (IBD) process,
\begin{equation}
    \antineutrino{e} + p^+ \to n + e^+. \label{eqn:IBD}
\end{equation}
The equivalent of this process for other flavours, for charge-conjugated variations, and for rotations of the process in flavour and charge space, are all of interest to various technology choices. We discuss some of these below.

\section{SNOLAB}
\label{sec:snolab}

\begin{figure}[ht!]
    \centering
    \includegraphics[width=\linewidth]{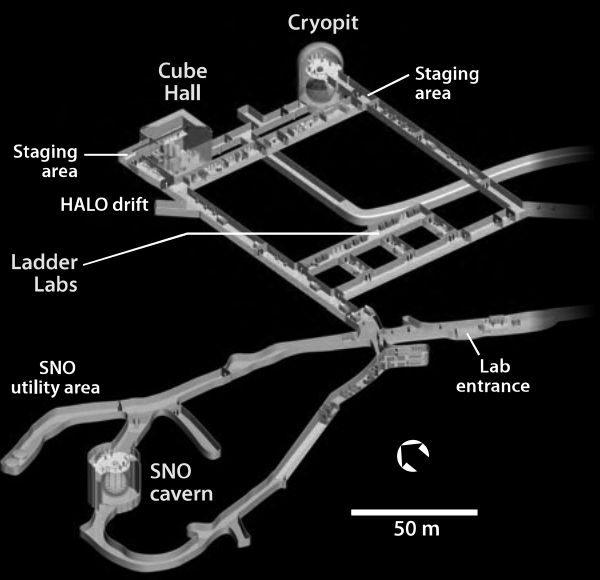}
    \caption{A map of the underground complex at SNOLAB. Figure adapted from Ref.~\cite{annurev.nucl.012809.104513}.}
    \label{fig:SNOLAB-map}
\end{figure}

SNOLAB is an underground cleanroom laboratory facility located at the Vale Creighton Mine in Lively, Ontario. The laboratory consists of a surface building (which hosts a range of spaces, including a cleanroom laboratory of similar quality to the underground facility) and the underground laboratory, located at the 6800~ft level (2.070~km) of the active nickel mine. The depth provides the laboratory with 6000~mwe of flat overburden composed of Norite rock. The shielding results in a cosmic ray muon rate of $3.31 \times 10^{-6}~\mathrm{\mu \cdot s^{-1} \cdot m^{-2}}$. This is a reduction by $5 \times 10^7$ as compared to the surface. A single active shaft (vertical tunnel) provides access to the drifts (horizontal tunnels) underground.

SNOLAB's underground complex is operated entirely as a clean lab with fewer than 2000 particles greater than 0.5 micron in a cubic foot (class-2000). The complex is $5000~\mathrm{m^2}$ in area, consisting of three large primary experimental halls (the SNO cavern, the Cube Hall, and the Cryopit) as well as a range of smaller spaces for hosting equipment and projects. The surfaces of the drifts that were excavated to host the laboratory were coated in a spray-on concrete (Shotcrete) and painted. Most of these were hand-troweled to make the surface smoother and easier to paint and clean. The rock, concrete, and paint can be sources of ambient radiation (see below).

Figure~\ref{fig:SNOLAB-map} illustrates the underground laboratory and indicates the location of experiments we discuss below, including HALO (\Cref{sec:HALO}) and SNO+ (\Cref{sec:SNOp}), as well as the Cryopit. In addition to experiments whose mission includes the detection of supernova neutrinos, many other experiments incorporate water tanks that are or can be actively instrumented, adding to the overall supernova sensitivity in the lab. Those experiments are detailed in \Cref{sec:other}.

In addition to the low cosmic ray muon flux, other laboratory backgrounds have been assessed~\cite{SNOLAB2006,SNOLAB2016}. Radon, a by-product of natural radioactive decay of uranium and thorium in the surrounding rock, results in about $130~\mathrm{Bq/m^3}$, up to 4 times higher activity levels than at the surface~\cite{Peterson2013-zt}. To compensate for this elevated level, some spaces within SNOLAB are operated as low-radon environments to prevent the ingress and absorption of radon onto instrumentation or materials. While radon itself decays over timescales short compared to experiment life cycles, the decay chain produces \Isotope{Pb}{210} with a half-life of 22~years. 

Spontaneous nuclear fission, $(\alpha,n)$, and $(\gamma,n)$ reactions in the surrounding rock and concrete produce neutrons, as do occasional interactions of cosmic ray muons with materials in and around the laboratory (muon spallation). The thermal neutron background has been assessed to be about $4.1 \times 10^{3}~\mathrm{n\cdot m^{-2} \cdot day^{-1}}$ with a 2\% uncertainty~\cite{Browne:1999pe}. The fast neutron background has been assessed to be comparable to the thermal flux but with a larger uncertainty. Finally, gamma radiation from nuclear processes in the rock and concrete contribute about $510~\mathrm{\gamma \cdot m^{-2} \cdot day^{-1}}$ for $E_{\gamma} = [4.5\textrm{--}5.0]~\mathrm{MeV}$, $360~\mathrm{\gamma \cdot m^{-2} \cdot day^{-1}}$ for $E_{\gamma} = [5.0\textrm{--}7.0]~\mathrm{MeV}$, and about $180~\mathrm{\gamma \cdot m^{-2} \cdot day^{-1}}$ for $E_{\gamma} > 7.0]~\mathrm{MeV}$. The uncertainties on these measurements are at the 50\% level, but gamma ray rates in the underground lab are $\mathcal{O}(10^3\textrm{--}10^4)$ times smaller than those at the surface owing to the reduction in cosmic ray shower activity.

\section{HALO}
\label{sec:HALO}

\subsection{Detector overview}

\begin{figure}[th]
    \centering
    \includegraphics[width=0.85\linewidth]{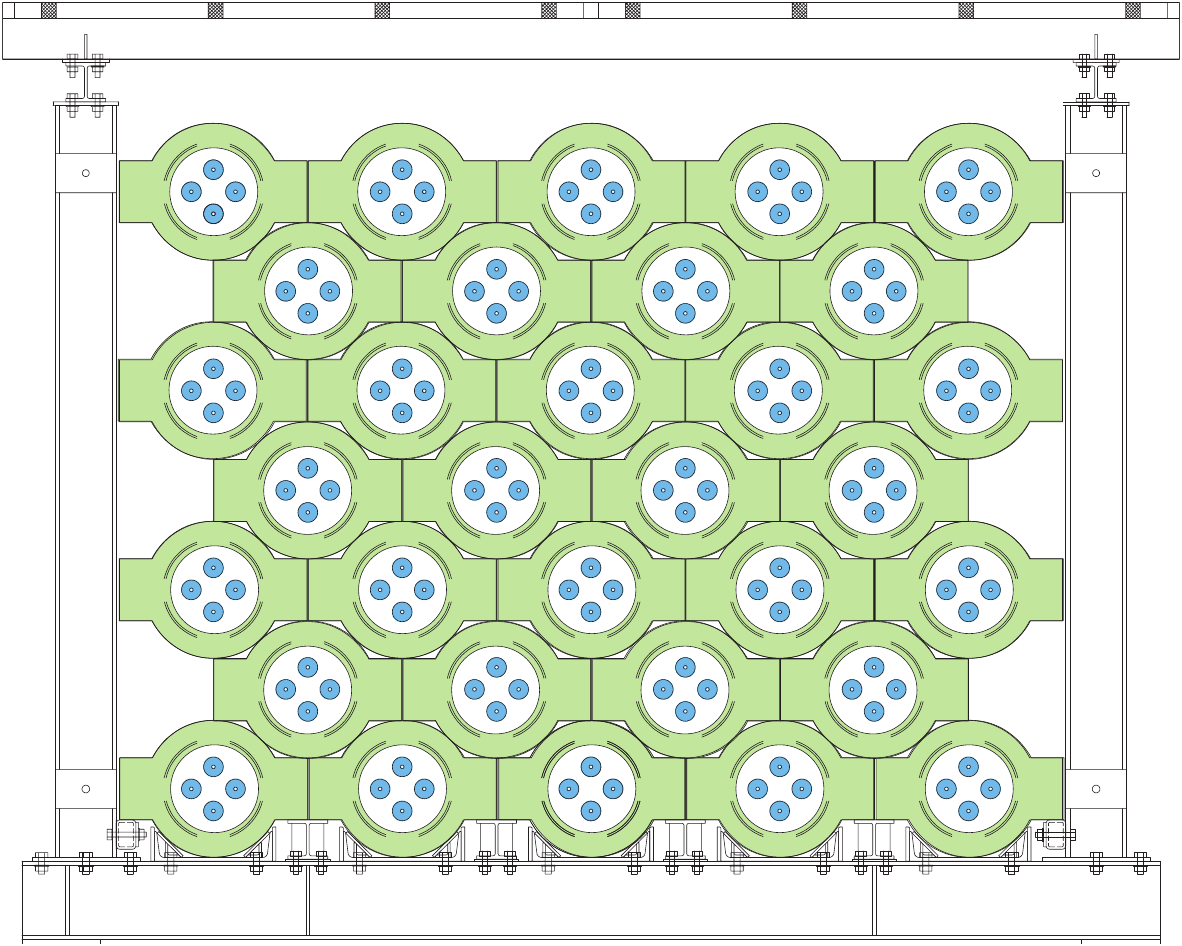}
    \caption{A schematic of the HALO experiment, showing the annular lead bricks (green-shaded semi-circles) and the \Isotope{He}{3} NCDs (blue-shaded circles). Figure reused with permission from Ref.~\cite{Vasel2014}.}
    \label{fig:halo_design}
\end{figure}

HALO (``Helium And Lead Observatory'') is a detector of opportunity, assembled at low cost from surplus materials from other physics experiments. Its design, construction, commissioning, calibration, and operation are detailed in a number of theses~\cite{Shantz2010,Bruulsema2017,Vasel2014,Hill2023}. The major components of the detector are lead, polyethylene-encased helium-3 proportional counters, and water boxes for external neutron shielding. There are in addition steel structural supports to ensure mechanical integrity.  The detector uses 79 metric tons of annular lead blocks that were transferred from a decommissioned cosmic ray station in Deep River, Ontario~\cite{Steljes1959}. Positioned inside the lead are 128 \Isotope{He}{3}-filled gas proportional counters that were built for the now-decommissioned SNO experiment~\cite{Amsbaugh:2007ke}. The latter are referred to as neutral current detectors in SNO and are repurposed in HALO as neutron counting detectors (NCDs). The HALO design concept is illustrated in Fig.~\ref{fig:halo_design}.  

The NCDs are placed in groups of four and read out in pairs: the top-bottom pair and the left-right pair. The pairs are read out by 64 pre-amplifiers and electrical signals are, in turn, input to eight analog-to-digital (ADC) converter boards, each with eight channels. These shaper ADCs can then be used to discriminate the signals from the NCDs using thresholds set in the system, and signals above threshold are digitized and saved. 

These digital signals are then sent to single-board computers running Linux. The data acquisition system is written in the ORCA framework~\cite{ORCA}. The detector is designed to operate inexpensively and with redundancy so that if any single component fails the experiment can continue to be ready for a supernova event.

\subsection{Detection principles}
The most abundant isotope of lead is \Isotope{Pb}{208}, which has 126 neutrons and 82 protons. Whereas a hydrogenous detector (organic scintillator like SNO+, or water Cherenkov like Super-Kamiokande) is primarily sensitive to $\antineutrino{e}$ interactions on the proton (see \Cref{sec:SNOp}), lead is sensitive to $\neutrino{e}$ (through CC and NC scattering) and $\neutrino{x}$ (through NC scattering).

The IBD reaction involves the conversion of a proton to a neutron. However, the lead nucleus hosts about 50\% more neutrons than protons, and the shell model of the lead nucleus favours this conversion only if there is a close available state in the neutron potential into which the converted proton could then go. Effectively, the large neutron-excess of the lead nucleus Pauli blocks the IBD process, and hence HALO is insensitive to charge-current interactions of electron anti-neutrinos. 

The favoured transitions are then of the type
\begin{equation}
 \neutrino{e} + n \rightarrow e^- + p.
\end{equation}
The conversion of a neutron to a proton is not Pauli-blocked, and thus a lead detector is primarily sensitive to $\neutrino{e}$.  The above nucleon-level process corresponds to the nuclear process
\begin{equation}
    \neutrino{e} + \Isotope{Pb}{208} \to e^- + \Isotope{Bi}{208}^*.
\end{equation}
The nuclear transitions populate Gamow-Teller (L=0) and electric dipole (L=1) excited states in the bismuth daughter nucleus, which decay predominantly by neutron emission. While the lead in HALO is dominated by \Isotope{Pb}{208}, it is by no means the only isotope present. We can generically denote the overall interaction and de-excitation process as
\begin{equation}
    \neutrino{e} + \Isotope{Pb}{\mathrm{A}} \to e^- + \Isotope{Bi}{\mathrm{A}}^* (\to  \Isotope{Bi}{\mathrm{A-1,2}} + (1,2)n + \gamma).
\end{equation}

\begin{figure}[th!]
    \centering
    \includegraphics[width=\linewidth]{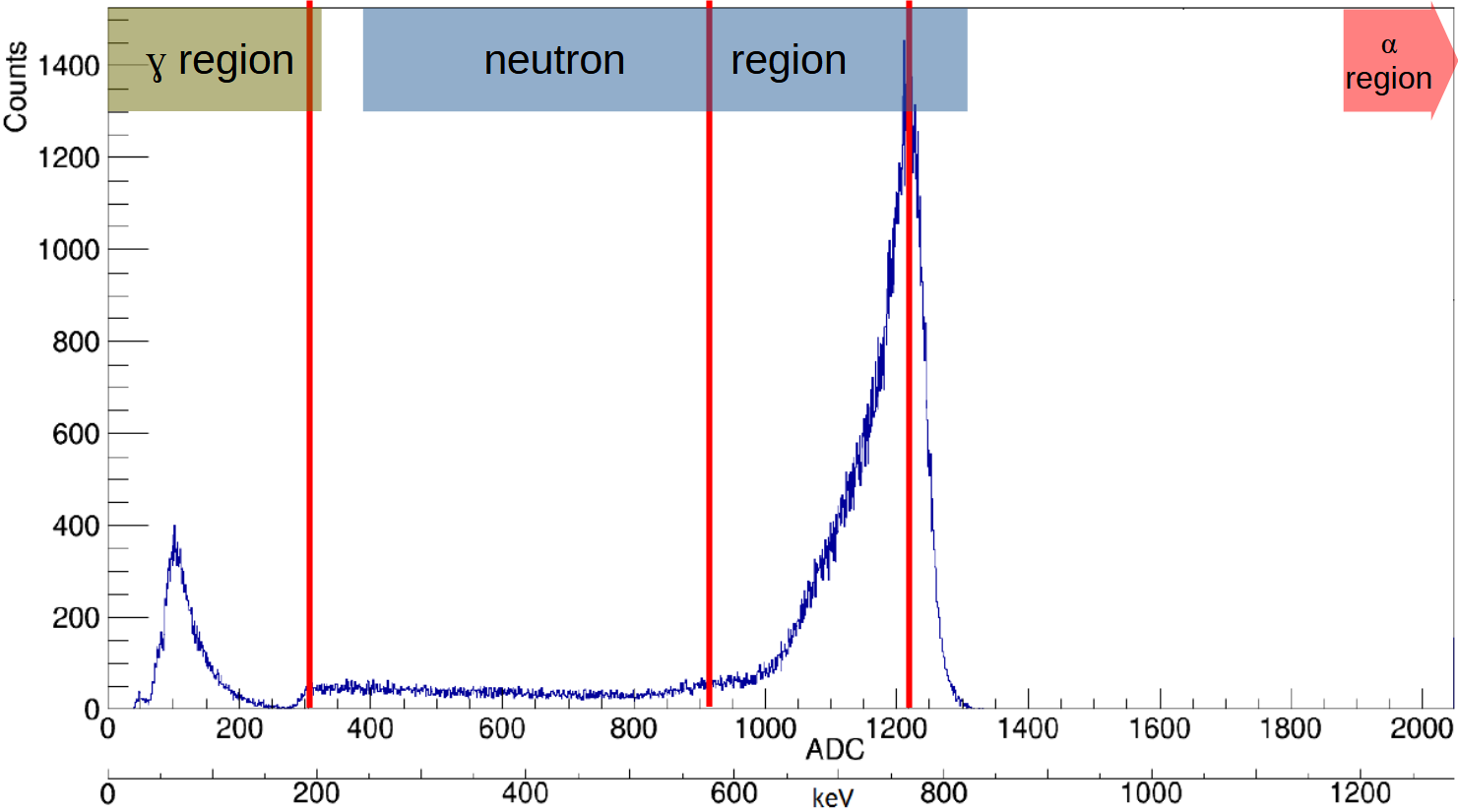}
    \caption{The energy spectrum observed by HALO during an 80-minute neutron calibration run with \Isotope{Cf}{252}. The most prominent peak corresponds to the complete collection of proton and neutron energy at 764~keV (or 1223 raw ADC), while the threshold corresponding to minimal energy detection (seeing only the triton) is visible at 191~keV. The proton-only observation threshold is visible starting at 573~keV. The structure below 300 ADC corresponds to gamma rays (``$\gamma$ region'') emitted by the walls of the NCD then Compton scattering in the gas volume. Alpha particles deposit large amounts of energy and would appear off the right-hand side of this graph (``$\alpha$ region''). Figure modified with permission from Ref.~\cite{Bruulsema2017}.}
    \label{fig:HALO_energy}
\end{figure}

The above CC process is dominant in lead, but the NC interaction still accounts for about 24\% of the neutrino interactions via the process
\begin{equation}
    \neutrino{i} + \Isotope{Pb}{\mathrm{A}} \to \neutrino{i} + \Isotope{Pb}{\mathrm{A}}^* (\to  \Isotope{Pb}{\mathrm{A-1,2}} + (1,2)n + \gamma)
\end{equation}
where $i$ runs over all three flavours. HALO is insensitive to CC anti-neutrino interactions. 

The lead-neutrino scattering cross-section is a key input to estimating the ultimate sensitivity of a detector like HALO. There have been multiple efforts to calculate this cross-section (c.f.~\cite{Kolbe:2000np,Engel:2002hg,Lazauskas:2007bs,Almosly:2016mse,Almosly:2019han,Ejiri:2019ezh}). 
If we consider a supernova neutrino energy of 15~MeV, the CC (NC) $\nu-\Isotope{Pb}{208}$ cross-section with single neutron production in the final state is at the level of $(1\text{--}2) \times 10^{-40}~\mathrm{cm^2}$ ($0.05 \times 10^{-40}~\mathrm{cm^2}$). The two-neutron channel is available at higher energies, and for $E_{\nu}=25~\mathrm{MeV}$ the overall CC (NC) cross-section is expected to be about $(10\text{--}17) \times 10^{-40}~\mathrm{cm^2}$ $((0.5\text{--}0.7) \times 10^{-40}~\mathrm{cm^2})$. These channels are denoted 1n and 2n, respectively. The uncertainties on theoretical estimates are generally not stated. HALO has proposed to measure the ratio of 1n-to-2n events during a CCSN burst as a means to probe the energy spectrum of the incident neutrinos.

A single experimental measurement of the lead-neutrino scattering cross-section exists and was made in 2023 by the COHERENT  Collaboration~\cite{COHERENT:2022fic}. Using a beam of neutrinos from a stopped-pion source at Oak Ridge National Laboratory, they report their observed cross-section in terms of a signal strength, the ratio of the  measurement and a prediction from the \textsc{MARLEY} event generator~\cite{steven_gardiner_2021_3905443,Gardiner:2021qfr}. Their predicted cross-sections were $42.1^{+4.7}_{-4.7} \times 10^{-40}~\mathrm{cm^2}$ assuming the 1n, 2n, and 3n final states and $39.3^{+4.7}_{-4.7} \times 10^{-40}~\mathrm{cm^2}$ for just the 1n and 2n final states. This prediction includes their assumed neutrino beam energy spectrum~\cite{doi:10.1126/science.aao0990}, which is in the range of about 16--53~MeV. COHERENT reports an observed lead-neutrino signal strength of $0.29^{+0.17}_{-0.16}$, relative to the prediction. Their dominant systematic uncertainty arises from the neutrino beam flux, accounting for 93\% of their total systematic error, $\pm 10.8\%$. Their total uncertainty is dominated by statistical effects. The uncertainty on the \textsc{MARLEY} prediction was estimated to be about $17\%$. There has been no clear determination of the possible cause of this suppressed signal strength, and as a result the implications to HALO are not clear, though possible effects are discussed later. 

Neutrino interactions will produce one or more neutrons that then traverse the HALO detector. The dominant lead isotope is ``doubly magic'', meaning that its proton and neutron shells are both full, and thus the neutron absorption cross section is small. Those neutrons will, nevertheless, thermalize by nuclear collisions in the lead matrix and especially in the 7.6~mm thick polyethylene moderator around the NCDs. Thermalization happens over a short timescale (around 20~$\mathrm{\mu s}$) while capture by helium, after thermalization, is a longer-timescale process (about $200~\mathrm{\mu s}$). These values were determined using calibration data, discussed below. 

Neutrons that enter the NCDs are very efficiently captured by the \Isotope{He}{3} nuclei. The detection reaction is
\begin{equation}
    \Isotope{He}{3} + n \to p^+ + \Isotope{H}{3} + \mathrm{764~keV}.
\end{equation}
Thus it is expected that each neutron interaction in an NCD will deposit at most 764~keV of energy, shared as kinetic energy between the proton (573~keV) and the triton (191~keV).  With the added 15\% of $\mathrm{CF_4}$ in the gas mixture, the triton has a range of only 1~mm in the NCD gas volume, while the proton has a range of 5~mm. If the neutron interaction occurs close to one of the NCD walls there can be incomplete collection of energy. If the proton (triton) exits the tube before completing its ionization of the NCD gas volume, HALO will observe energies between 191--764~keV (573--764~keV). The energy spectrum from a neutron calibration (using a \Isotope{Cf}{252} neutron source) is shown in Fig.~\ref{fig:HALO_energy}.

\subsection{Detection backgrounds}

The ``neutron region'' of the energy spectrum is generally defined as being between 380--1300~ADC. A simulation of the HALO detector, developed in GEANT4, was used in conjunction with calibration data to assess the neutron efficiency of the detector. A \Isotope{Cf}{252} source was positioned in 192 different locations in the detector using a series of 24 source tubes embedded in the lead matrix, and the source was kept at each position for 9~minutes. The activity level of the source, combined with knowledge of its neutron multiplicity spectrum, was used to determine the neutron efficiency as a function of position in the HALO detector. The per-neutron efficiency is almost 50\% near the centre of the detector and 10\% near the edge, with an average of 27.6\%. The simulation is calibrated using the \Isotope{Cf}{252} data. The neutron detection efficiency for a simulated supernova burst signature is estimated to be 28.3\%. Thermalization and capture timescales were measured in decays with multiple neutrons, using the earliest neutron capture interaction as $t=0$ for determining the timescales of the other neutrons. Neutrons travel, on average, about 50--60~cm from their production point before being captured.

Gamma rays do not significantly populate the neutron region as they interact via Compton scattering and the subsequent electrons have energies typically below 200~keV. However, this energy region is useful in the event of a claimed supernova neutrino burst as a cross-check of the detector's performance stability. Alpha particles emitted from the walls of the NCDs are detected, but they deposit a large amount of energy through ionization and have ADCs in excess of 2000. 

Electrical noise can also be a source of background in HALO and happens a few times per month. It can be caused by seismic activity (shaking the electronics) or any other vibrational event near HALO. This generates a large number of counts across the NCDs but at very low energies, and is readily distinguished from neutrons.

The primary backgrounds are then from real neutrons, either from the surrounding laboratory materials, or induced in the detector by spontaneous fission (SF) or from muon spallation. Neutrons from surrounding materials would be expected to result in random detection coincidences in the NCDs. These primarily contribute to a background rate of 1-neutron events, but can through coincidence also result in $N$-neutron bursts (where $N>1$). This ambient Poisson neutron background is estimated to result in about 1200 neutrons per day.

Neutrons from SF would be expected to have low multiplicities (in the range of 1--3 neutrons per SF) and a time structure that is very short, since neutrons are produced all at once from a single event and then are captured after thermalizing. A study of $N=[2,3,4]$-neutron bursts (where a burst is defined at $N$ neutron detections within a 1.5~ms time window) found that the burst rates from SF were estimated to be 3.5 ($N \geq 2$), 0.8 ($N \geq 3$), and 0.2 ($N \geq 4$) per day. These estimates are made after subtracting the random coincidence background, estimated using the laboratory neutron background rates. Taking into account the neutron multiplicities, this results in a total estimate of about 10 neutrons per day from SF.

Neutrons from muon spallation will result in higher-multiplicity neutron bursts as the muons inelastically scatter off nuclei around the detector or in the lead. Bursts above a multiplicity of $N=4$ and within a 10~ms window are classified as ``spallation''. After subtracting off the extrapolation SF burst rate for $N \ge 4$, it is estimated that HALO will observe 0.07 spallation events per day. Spallation bursts were found to have overall time structures that were typically shorter than 3.5~ms. Taking into account the neutron multiplicities, the total background estimate from spallation is about 1 neutron per day.

\subsection{Detection sensitivity}

A supernova neutrino burst would be expected to produce multiple interactions in a 10-second window. 
Taking into account HALO's efficiency and the backgrounds above, the supernova burst trigger is set to a threshold of at least four neutrons in a 2-second window. 
This is estimated to result in about four false positive supernova alerts each year, below the acceptable maximum of six per year defined by the SNEWS collaboration.

Supernova simulations are further used to estimate the trigger efficiency as a function of the distance to the supernova. The sensitivity is dependent on the peak of the neutrino energy spectrum. 
The calculations of ~\cite{Vaananen:2011bf} provide an estimate. The theoretical $\nu\text{--}\mathrm{Pb}$ interaction cross-section of ~\cite{Engel:2002hg} ~is used instead of instead of the COHERENT experimental result.  Taking initial average energies of 10, 13 and 18~MeV for $\neutrino{e}$, $\antineutrino{e}$, and $\neutrino{X}$, respectively,  and the pinching parameter $\eta_{\alpha}=2$, the lead matrix of HALO is expected to release 55 neutrons, half within the first 2 seconds.  
With the threshold of four neutrons within a 2-second window, HALO is expected to be $>68\%$ trigger efficient for supernovas within $13.7~\mathrm{kpc}$, and $>95\%$ efficient within $10~\mathrm{kpc}$~\cite{Bruulsema2017}. %
If the cross-section is 29\% of expected (as measured by COHERENT), then the distances would decrease by a factor of 54\%, and supernova sensitivity distances would change to $7.4$, and $5.4~\mathrm{kpc}$, respectively.

\section{SNO+}
\label{sec:SNOp}

\subsection{Detector overview}

\begin{figure}
    \centering
    \includegraphics[width=0.8\linewidth]{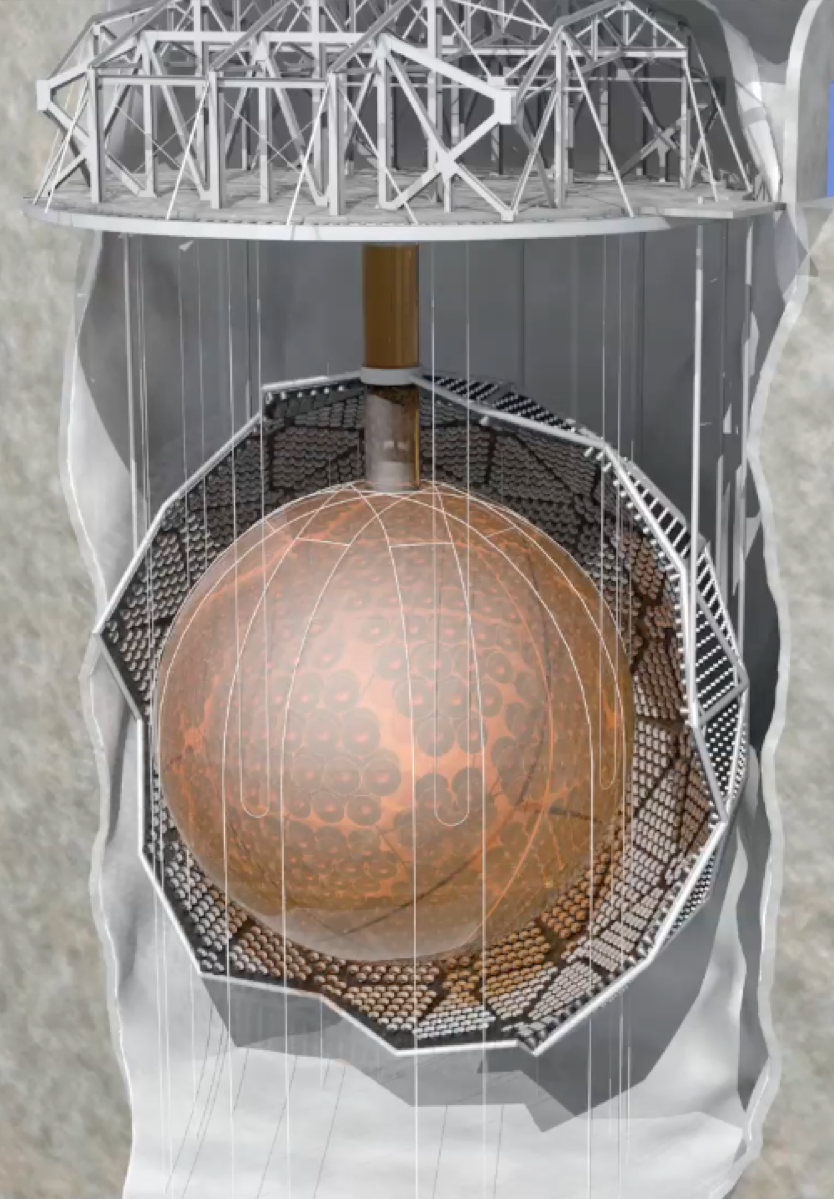}
    \caption{An artist's depiction of the SNO+ detector, Figure reused with permission from Ref.~\cite{SNOP_Detector}. \copyright~IOP Publishing Ltd and Sissa Medialab. Reproduced by permission of IOP Publishing. All rights reserved.}
    \label{fig:SNOP_detector}
\end{figure}

SNO+ is a large neutrino detector located at SNOLAB and occupies the single-largest cavern in the facility. Much of the hardware was reused from the SNO experiment~\cite{SNO_detector}. The main target volume is contained within an acrylic vessel (AV) that is $12~\mathrm{m}$ in diameter but only $5~\mathrm{cm}$ thick. This is suspended from a deck that is at the top of the SNO cavern and affixed to the cavern walls. The AV is secured by a rope net that attaches to the cavern floor. The cavern is filled with 5300~tonnes of ultra pure water (UPW). Surrounding the acrylic vessel is a structure to hold $\sim 9300$ photomultiplier tubes (PMTs). Between the PMTs and the acrylic vessel is another 1700~tonnes of UPW. The PMTs provide $\sim 50\%$ photo-coverage. More details about the detector construction are found in Ref.~\cite{SNOP_Detector}, and a depiction of the experiment is shown in \Cref{fig:SNOP_detector}. 

The signals from the PMTs are read out by custom electronics.
The charge and timing of each triggered PMT are recorded and built into events for analysis.

The AV is filled with a scintillator cocktail: $780$~tonnes of linear alkylbenzene (LAB), $2.2$~g/L of the wavelength shifter 2,5-Diphenyloxazole (PPO), $2.2$~g/L of the fluor 1,4-Bis(2-methylstyryl)benzene (bis-MSB), and $1.1$~g/L of the stablizer butylated hydroxytoluene (BHT). The final phase of SNO+ will introduce 1300~kg of \Isotope{Te}{130}\ for studying neutrinoless double beta decay.

\subsection{Detector sensitivities}
 
\begin{figure}
    \centering
    \includegraphics[width=\linewidth]{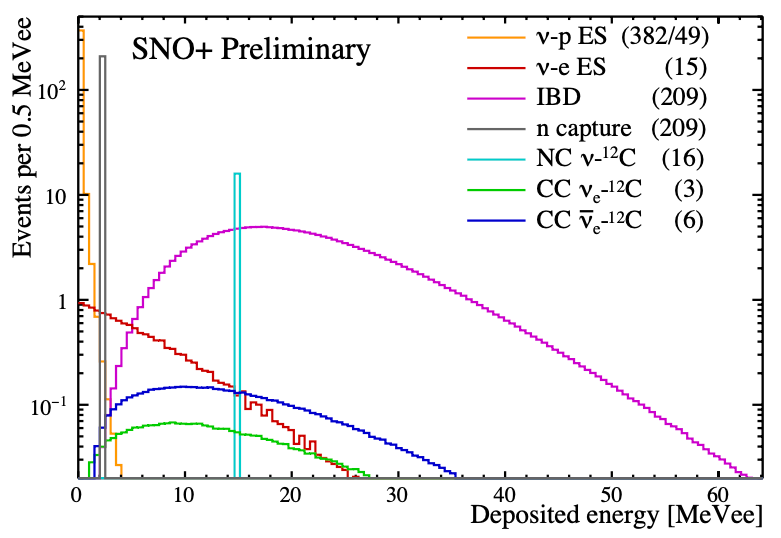}
    \caption{The simulated raw energy spectrum (electron equivalent energy) for events from a supernova at 10~kpc using the LS220-s27.0co model \cite{mirizzi_2016}. The legend indicates the expected number of events for each signal, with the second number for the $\nu\text{--}p$ proton recoil events indicating those events above a 200~keV threshold. Figure reused with permission from Ref.~\cite{Rumleskie2021}.}
    \label{fig:SN_nu_en_spectra}
\end{figure}

A supernova generates a burst of high-energy neutrinos, which may be detectable with detectors like SNO+. 
SNO+ will be sensitive to CCSN neutrino bursts throughout all phases of data taking, with a threshold of around 200~\keV (limited by \Isotope{C}{14}\ beta decays) allowing a rich range of signals potentially including neutral-current proton recoil events that detect all neutrino flavours. 
\Cref{fig:SN_nu_en_spectra} shows the predicted raw energy spectra of events from a CCSN at 10~kpc simulated using Ref.~\cite{huedepohl_2010}. 

In addition to interactions in the scintillator, there will also be neutrino interactions in the external water regions  generating many IBD events and some elastic-scattering Cherenkov events with pointing information to the parent supernova (for a relatively nearby CCSN that could have a sufficient $\neutrino{e}$ event rate). 
Directionality for elastic-scattering events in LAB, as already observed in SNO+ with low PPO concentration~\cite{SP-Directionality}, may also be possible in the final LAB deployment for the higher energy CCSN neutrinos. 

The IBD reaction is the preferred method for studying electron antineutrinos in organic scintillator because of its high cross section, the abundance of protons in the target, and the high energies it produces. 
As a CC process, both the neutron and positron products from Eqn.~\ref{eqn:IBD} emit visible light in the liquid scintillator. 
The positron quickly deposits its energy and annihilates, releasing two $511~\keV$ photons, which appear as a prompt signal in the detector. 
Approximately $200~\mus$ later, the neutron is captured on hydrogen,
\begin{equation}
    n + p \rightarrow d + \gamma~(2.223~\MeV)
\end{equation}
creating what is known as the \emph{delayed event}. 
By analyzing the timing and spatial locations of these prompt and delayed events, the IBD reaction can be identified with high efficiency, significantly reducing background processes. 
The threshold for this process is $1.8$~\MeV, which is low compared to the expected CCSN energy of $\sim 10~\MeV$.

Also available is the proton scattering reaction, $\nu + p^+ \rightarrow \nu + p^+$. 
Proton scattering is a purely neutral-current process that occurs equally for the different neutrino flavours. 
The cross section for proton scattering and its differential form may be used to determine the neutrino and proton recoil energies, assuming the incoming and outgoing angles are known with sufficient precision.  
The energy response of a liquid scintillator detector is non-linear due to energy quenching of the scintillator. 
The proton energy will be quenched, and light output of the scintillator is reduced. 
The energy response of the detector for proton scattering is currently under investigation. 

A neutrino can scatter elastically off an electron in the detection medium via  $\nu+e^{-} \rightarrow \nu+e^{-}$. The electron’s energy, $E_e$, after the scattering can be calculated using the conservation of energy and momentum by assuming the electron is initially at rest. Neutrinos of all flavours can scatter off electrons, and the maximum energy of the scattered electron is given by 
\begin{equation}
    E_e^m=\frac{2E_{\nu}^2}{m_e+2E_{\nu}}.
\end{equation}

There are two visible interactions on carbon that occur through the CC interaction, one with the electron neutrino:
\begin{equation}
\neutrino{e} + \Isotope{C}{12} \rightarrow \Isotope{N}{12} + e^{-}     
\end{equation}
and one with the electron antineutrino:
\begin{equation}
\antineutrino{e} + \Isotope{C}{12} \rightarrow \Isotope{B}{12} + e^+.     
\end{equation}
These reactions have thresholds of 17.86~\MeV\ and 13.89~\MeV, respectively, and can produce a nucleus either in its ground or an excited state. The final-state charged lepton creates prompt visible light in the detector. Additionally, the resulting nucleus from the reaction will decay, producing detectable products that appear as delayed events in the detector.

Finally, neutrinos and antineutrinos may also interact with \Isotope{C}{12} via the NC process. 
The interaction leaves the carbon nucleus in an excited state that emits a 15.1~\MeV\ gamma ray upon de-excitation:
\begin{align}
   \nu + \Isotope{C}{12} \to &\Isotope{C}{12}^* + \nu     \\
    &\Isotope{C}{12}^* \to  \Isotope{C}{12} + \gamma~(15.1~\MeV)
\end{align}
More events are expected from this interaction (which can occur for all neutrino types) than from both CC interactions on carbon. 
During a supernova, the 15.1~\MeV\ gamma ray is expected to be visible above the prompt energy spectrum from IBD reactions. All of these reactions are summarized in \Cref{tab:SN_in_SNO+}.

{
\renewcommand{\arraystretch}{1.5}
\begin{table}[h!]
\centering
\begin{tabular}{|p{4cm}|l|}
\hline
\textbf{Reaction } & \textbf{Process} \\ \hline %
Inverse Beta Decay & $\antineutrino{e} + p \rightarrow e^+ + n$ \\ \hline%
Proton Scattering & $\nu + p \rightarrow \nu' + p$ \\ \hline%
Electron Scattering: & $\nu + e^- \rightarrow \nu' + e^-$ \\ \hline%
CC:~Carbon Scattering & $\neutrino{e} + \,\Isotope{C}{12} \rightarrow \,\Isotope{N}{12}+ e^-$  \\ \hline%
CC:~Carbon Scattering & $\antineutrino{e} + \,\Isotope{C}{12} \rightarrow \,\Isotope{B}{12} + e^+$  \\ \hline%

NC:~Carbon Scattering & $\nu + \,\Isotope{C}{12} \rightarrow \,\Isotope{C}{12}^* + \nu'$ \\ 
& $\Isotope{C}{12}^* \rightarrow \,\Isotope{C}{12} + \gamma (15.1 \, \text{MeV})$ \\ \hline%

\end{tabular}
\caption{Detection reactions of supernova neutrinos in SNO+. See~\cite{Rumleskie2021, Rigan2021}.}
\label{tab:SN_in_SNO+}
\end{table}
}

All reported sensitivity studies use the volume of the acrylic vessel and its target material. They do not include neutrino interactions in the $7$~kT of shielding water.
The sensitivities are calculated using the SNUGen software \cite{VonKrosigk:2015yio}. This code requires the average  energy, luminosity, and pinching parameter of each neutrino flavour (\neutrino{e}, \antineutrino{e}, \neutrino{x}, \antineutrino{x}). 
Input files of this data are provided by the Garching Group \cite{garching} for set of 1D models with progenitor masses ranging from 8.8 -- 40.0~\msolar. 

\subsection{Calibration Sources}

Simulations predict that the SNO+ detector will have a significant response to CCSN neutrinos, with light production at a level capable of saturating the data acquisition system. 
To calibrate the detector for such an event, a special method has been developed. 

The supernova calibration source mimics the sequences of light pulses produced by neutrino interactions within the detector by channeling pulsed light from a high-powered (120~mW), blue-violet (405~nm) laser diode into the an acrylic diffusing sphere, called a \emph{laserball}. 
The laser diode is mounted into a socket on a custom-designed printed circuit board hosting the circuitry for its operation.

Using pulse durations modeled on CCSN signals~\cite{mirizzi_2016, huedepohl_2010}, the source can emit bursts of photons, corresponding to a total deposited energy of up to 100~MeV within the scintillator. 
This results in all PMTs in the detector receiving multiple photons. This is an otherwise rare occurrence during normal operations, leading to a low false-positive rate. 
Testing the data acquisition system this way assesses the detector’s capacity to capture and record data from a nearby CCSN, where numerous neutrino interactions occur in quick succession.

To replicate a realistic supernova neutrino burst, each event is defined by pulse energy and timing separation between pulses from the laser diode. 
The expected supernova energy distribution is sampled for pulse energy, while a Poisson distribution related to the neutrino interaction rate at the event time determines the timing separation between pulses. 
These pulse energy and timing values are encoded as 32~bits, and all events for each supernova burst are stored in a single ``burst'' file. 
Burst files are pre-generated based on various supernova models, neutrino oscillation scenarios, and distances~\cite{Darrach2016}.

An FPGA reads these burst files, translating them into signals for the laser diode driver board. Timing information sets a countdown timer running on a 12.5~MHz clock, allowing for realistic event pile-up within the 400~ns trigger window. 
Energy information is mapped through specific look-up tables to a current, which powers the laser diode after the countdown completes. 
Light from the laser diode is then transmitted through an optical fibre bundle to the laserball, which disperses it isotropically from the centre of the AV. 
A shutter, normally closed to prevent laser exposure to the detector and operators, sits between the laser diode and fibre bundle and only opens during calibration runs when the system is confirmed safe.

Another method to simulate a CCSN-like event in the SNO+ detector uses an existing \emph{in-situ} optical calibration system ~\cite{SNOP_Tellie}. 
The Embedded LED/Laser Light Injection Entity (ELLIE) is used to shine broad beams of light through optical fibres across the AV at a set frequency, intensity, and duration.
The frequencies and intensities can be modified for a series of runs to simulate the expected signal from a supernova
This system was used to determine the response of the detector readout system under such conditions.
Three fibres were chosen, and each was run with increasing frequency of light pulses until data-loss was observed. 
Once the data acquisition limits were known, the distance of an expected supernova spectra can be compared to determine the detector's sensitivity, assuming that the test accurately simulates a supernova signal. 
From the test, a full saturation of the detector is estimated to occur from an event between 1.4--2.4~kpc~\cite{Rigan2021}.

\subsection{Burst Trigger}

The characteristic CCSN neutrino burst is used to develop a burst trigger focusing on high-energy events.
The SNO+ trigger thresholds for energy and event multiplicity have been re-evaluated. 
SNO+ can adjust the lower energy ranges to maximize its physics output while satisfying any external tolerance for false positives. 
Including lower energy events may reveal signals from proton elastic scattering events, which SNO+ is uniquely positioned to detect among global neutrino detectors.

The SNO+ burst trigger operates through three levels~\cite{Rigan2021}. Level 1 involves the data acquisition and event triggering system. 
In Level 2, the process begins with a burst declaration time window of up to 2~s, during which events are evaluated against thresholds to determine if a burst should be declared. 
Once these thresholds are met, a burst is declared, and an ``extending window'' of one second begins. 
If an event within this extending window has a hit multiplicity exceeding 700~PMTs, the window renews for an additional one second from that event, with a maximum extension of up to 42~s. 
Together, the declaration and extending windows form the primary burst detection period. 
There is an additional ``IBD Burst Channel,'' which tags two IBD events within 2~s and is not dependent on the total number of PMTs hit, unlike the other trigger channels.
Additionally, the logic includes pre-burst and post-burst buffers to monitor event trends, recording all events 1~s before the declaration window starts and 1~s after the extending window ends.

Thresholds for the Level 2 burst monitor are calibrated to maintain a false burst rate of approximately once per month. Backgrounds arise primarily from radioactive decays~\cite{Wang2022}. 
The burst file then advances to Level 3 for data cleaning, where the cleaned events are re-evaluated by the burst logic. 
If they still meet burst criteria, the Level 3 burst is flagged for manual review. 
In the future, the SNO+ supernova team plans to automate notifications to the SNEWS network for Level 3 burst events. SNO+ is preparing to join SNEWS2.0 (see \Cref{sec:snews}). 
SNO+ is already sending test alarms to the system and is preparing to send automated alerts.

\subsection{Pre-Supernova Neutrinos}
Pre-supernova neutrinos are the neutrinos emitted from pair-annihilation cooling in massive stars during the silicon burning phase, prior to becoming a supernova~\cite{KamLAND:2015dbn}. 
If pre-supernova neutrinos are detected, they would signal an imminent CCSN a few days before the core collapse neutrino burst event. SNO+ operates a pre-supernova trigger that selects IBD events and creates an alert for a $3\sigma$ excess above expected rates within a 12~h period. 
The sensitivity, parameterized in terms of the time of warning before collapse, has been studied for a nominal IBD interaction rate from reactor- and geo-antineutrinos and SNO+'s measured $(\alpha,n)$ rates. There is expected sensitivity for 15--25~\msolar\ progenitors that are very close (within $300$~pc), under all flavour oscillation scenarios. 
SNO+ plans to share its pre-supernova alert in a manner similar to the combined KamLAND and Super-K system~\cite{combo_preSN}. 

\section{Other SNOLAB Experiments}
\label{sec:other}

The SNO+ and HALO experiments are extant projects whose primary research portfolio included the detection of supernova neutrinos. In addition to these, there are planned projects that either include supernova neutrinos as a primary physics goal or that have explored sensitivity to them even if they are not a physics goal. We briefly summarize a few of those here and indicate potential future opportunities at the SNOLAB facility.

\subsection{PICO-500}
\label{sec:pico}

The PICO-500 experiment~\cite{Garcia-Viltres:2021swf} is a bubble chamber designed to detect low-energy nuclear recoils. Its primary physics goal is the detection of dark matter with an energy threshold at the level of 1--10~keV that can be tuned using the temperature and pressure of the target liquid. The liquid is shielded by a surrounding water tank, stainless steel pressure vessel, and mineral oil buffer so that it operates in a very low-background environment. The threshold for bubble nucleation in the fluid is tuned by changing the temperature of and pressure exerted on the fluid. The goal is to keep the liquid in a superheated state so that when any process deposits energy in a small volume and above the selected threshold, boiling initiates in a localized zone.

PICO-500 is under construction at SNOLAB and will host a 260~L octafluoropropane ($\mathrm{C_3 F_8}$, or freon) target with a mass of about 270~kg. Since any nuclear interaction that deposits energy above the operating bubble nucleation threshold will induce the formation of a bubble, the experiment is sensitive to any particle that can penetrate into the target volume and induce such a reaction. Common backgrounds arise from alpha particles (radioisotope decay) and neutrons (spallation). 

A range of optical and acoustic techniques are used to distinguish backgrounds from dark-matter-like single-site nuclear recoils (expected to induce single bubbles in the volume). Neutrons will multiple-scatter, leading to ``multibubble'' events that are distinct from single-bubble events. The rate of neutron-induced single-bubble events can be inferred from the multi-bubble population. All other backgrounds, as well as a dark matter signal, should induce single-bubble events.

Neutrinos can also induce bubbles through interactions with the freon target nuclei. PICO-500 is expected to be sensitive to the solar neutrino background within a few years of operation, depending on the operation threshold. Sensitivity to CCSN neutrinos has been explored using assuming coherent elastic neutrino scattering (\cevns ~on the freon target)~\cite{Kozynets:2018dfo}. 
 
The study determined that the expected topology of a supernova neutrino event was a multi-bubble trigger induced by multiple neutrino interactions in the volume in a short period of time. The assumption was that the experiment's shielding would reduce neutron multi-bubble events to an acceptable false-positive level.

The study employed a pinched CCSN supernova model with the progenitor event occurring at various distances from Earth and a post-bounce time of 16.8~s. Sensitivity is quoted for a CCN at 10~kpc. The neutrino spectrum peaked around 12~MeV for a progenitor of 20~\msolar. They estimated that for energy thresholds between 2--10~keV, PICO-500 with 750~L of target fluid would see 2--4 bubbles over the post-bounce time period. At the minimum energy threshold, assumed to be 2~keV, such a target would expect to observe 2 bubbles within 2 seconds of the arrival time of the neutrinos, and 3 within 5 seconds.

However, all of the above assumed the initially designed volume of 750~L of freon. After successive iterations of design evolution, the final detector will use almost a third of that, 250~L. A down-scaling of the above conclusions to a realistic freon target implies the actual sensitivity of PICO-500 would be significantly reduced.

We note that the authors of the original study did not factor in other possibilities, such as using the water tank surrounding PICO-500 either as its own neutrino detector or in coincidence with the freon volume in a combined trigger. The water tank is expected to hold about 174,000~L of ultra-pure water and to be instrumented with an array of 48~PMTs for use as a cosmic ray muon veto~\cite{HawleyHerrera:2024wlg}. This will help to tag and reject the spallation neutron events in PICO-500, but could also be operated in a trigger configuration where a multi-PMT detection of Cherenkov radiation (\'a la SNO+), in coincidence with one or more bubbles in the freon target (over the time window) could be used to construct a supernova trigger.

\subsection{nEXO}
\label{sec:nexo}

The nEXO Experiment is a planned next-generation neutrinoless double beta decay detector~\cite{nEXO:2021ujk}. It will use an active volume containing about  5000~kg of xenon enriched to the 90\% level in \Isotope{Xe}{136}. This volume is enclosed in a time projection chamber (TPC) that itself sits within a series of nested vessels, all of which is mounted inside a large water shield, referred to as the outer detector (OD). The water shield will be instrumented with PMTs for use as a muon veto. 

Both the xenon TPC~\cite{nEXO:2024uyw} and the OD~\cite{Kharusi:Phd:2024} have been explored as supernova detectors.
Estimates have used baseline sets of CCSN neutrino flux models (GVKM~\cite{PhysRevLett.103.071101}, Livermore~\cite{totani1998future}, and pinched-thermal~\cite{Minakata:2008nc,Tamborra:2012ac} models). 

Detection is expected to be dominated by CC scattering. In the TPC it would be through $\neutrino{e} + \mathrm{{}^{134,136}Xe} \to e^{-} + \mathrm{{}^{134,136}Cs^{*}}$ and in the OD by IBD. Neutral current processes are at least $10\text{--}100$ times smaller in each system but have been considered. Recent work has focused on using these systems independently, though there is potential for a combined trigger.

nEXO expects to observe neutrinos from CCSN as far as $5\text{--}8~\mathrm{kpc}$ ($10\text{--}20~\mathrm{kpc}$) via TPC-only (OD-only) methods. A burst algorithm in the OD has been proposed, using a 5-fold PMT coincidence at the single photoelectron level with a window of 240~ns. Studies have so far ignored detector backgrounds and draw conclusions using the GVKM model. Additional future work on this subject from the nEXO collaboration is anticipated.

\section{Additional Project Resources at SNOLAB}
\label{sec:additional}

We note that SNOLAB provides large water shields for experiments in the form of cylindrical, welded stainless steel, grain silo tanks. There are currently four of these, in addition to the SNO+ cavern and the anticipated renovation of the cryopit for a next-generation large-scale experiment~\cite{SNOLABImplPlan2023.2029}. Their dimensions (diameter $\times$ height) are as follows: 7.9~m$\times$7.8~m or 377,000~L (DEAP-3600), 5.6~m$\times$7.9~m or 195,000~L (PICO-500), 2.9~m$\times$3.7~m or 24,000~L (PICO-40L) and 3.7~m$\times$3.2~m or 35,000~L (CUTE). The combined volume of these tanks is 631,000~L. For comparison, the SNO+ acrylic vessel holds about 900,000~L of target fluid (the basis of their supernova sensitivity estimates). 

Each of the existing tanks is intended to hold an experiment that takes up non-negligible volume. Nevertheless, if instrumented (with PMTs, for example) these tanks could be combined to provide an additional neutrino-sensitive water Cherenkov target whose volume approaches at least half that of the SNO+ vessel. Experiments using these water tanks now or in the future might consider the value of adding supernova trigger algorithms to their instrumented water shield usage, and/or instrumenting their water shields for this purpose. These could then be joined in a SNOLAB-level observatory effort, or each individual trigger could be sent to SNEWS for a combined decision using that system. 
While the sum of the signals from different water tanks would not have the same efficiency or resolution as a single tank of the same volume, having a number of signals from one location would still be of value to the SNEWS sensitivity.

\section{SNEWS}
\label{sec:snews}

The Supernova Early Warning System (SNEWS)~\cite{Antonioli:2004zb,AlKarusi:2021} is a collaboration of scientists spanning neutrino experiments across the globe. The collaboration operates a central coincidence system (at two redundant sites) that receives alerts from individual experiments in the collaboration. The system has been operational since 1998, fully automated (SNEWS1.0) since 2005, and is being upgraded (SNEWS2.0). 

The coincidence system then decides, based on the number and timing of such alerts from those experiments, whether or not to issue a general alert to the astronomical community of a potential impending observable supernova. The original goals of SNEWS were to provide prompt alerts that were positive (meaning reducing the chance of a false positive) and provide some measure of pointing information. SNEWS2.0 reduces the burden on experiments to avoid the submission of false-positive triggers.

The SNEWS collaboration notes the important role of different technologies in resolving model differences with a future detection but makes no recommendations on baseline models that could be used to calibrate sensitivity across detectors. The infrequency of galactic supernova and the need to provide reliable alerts to the community drove the SNEWS1.0 design goal of issuing not more than one false-positive alert per century. To do this, they considered the probability of an n-fold false coincidence in 10-second windows when receiving false-positive alerts from independent experiments. As an example, a 2-fold false coincidence from five operating experiments is expected to happen about once per 100 years, whereas a 3-fold such coincidence would happen about once per few million years. SNEWS1.0 was not focused on pointing back to the CCSN in the sky, but SNEWS2.0 emphasizes that capability.

SNEWS classifies alerts as ``Gold'' and ``Silver''. If input experiments meet high standards of quality (e.g., low alert rates) and yield a coincidence within 10~s, the alert is ``Gold'' quality and is sent first to other experiments in the SNEWS network as well as gravitational wave observatories and gamma ray observatories. ``Silver'' alerts are shared only with other experiments in the SNEWS network.

The SNEWS coincidence system accepts data from each member experiment with the name of the experiment, an estimate of the time of the first event detected, and a quality parameter describing that experiments degree of confidence in the burst that triggered the alert. SNEWS estimates at 90\% confidence level the region of the sky corresponding to the location of the neutrino burst, determined using the geographical locations of the experiments and their first event times. The goal is to deliver a region to other observatories with an uncertainty at the level of a few degrees.

\section{Conclusions and Outlook}

The models for CCSN we have discussed in this paper suggest a range of possible outcomes for the neutrino population that ultimately reaches Earth. Models (including the mass hierarchy of the neutrinos) can lead to different predictions about the electron- and other-flavour neutrino populations, and anti-neutrino populations, that can be detected. The experiments we have highlighted here in Canada already provide complementary sensitivity to different flavour and charge-conjugation states for neutrinos, allowing for comparison the different experiments to infer information about the neutrino population.

SNO+ and HALO, along with the other experiments, will play an important role in supernova neutrino observation because of their location, size, and flavour sensitivity.
Even the location of SNOLAB is important, especially to the sky localization of the next CCSN. Though the detectors hosted at SNOLAB are currently smaller than those in other laboratories, SNOLAB's geographical distance from those labs will provide a critical ``lever arm'' for triangulation. We emphasize that the community's ability to study the next galactic CCSN benefits from more detectors spread across the globe, even if there are concentrations of approaches in one or more locations.

We have also noted other potential opportunities for SNOLAB. In addition to future experiments whose active detector volume should provide some sensitivity to CCSN neutrinos, the shielding of those projects can be made active to increase the target volume sensitive to neutrino interactions. This might also just include outfitting existing water infrastructure with active elements, independent of hosting an experiment in the water volume.

Canada currently is a global leader in hosting and supporting CCSN-sensitive experiments. We encourage a continuation of support for leadership in this area. Though a galactic CCSN is a rare event, the challenges that must be overcome to assure detection are substantial and more redundancy in infrastructure will guarantee we seize the next CCSN observation opportunity.

\begin{acknowledgments}
We are grateful to Sam Hedges, Hamish Robertson, and Jeff Tseng for input on this paper. We are grateful to Stephen Stankiewicz and Paul Larochelle at SNOLAB for their communication regarding the volumes of stainless steel water tanks in the laboratory. \\
The authors declare there are no competing interests.
The authors declare no specific funding for this work.
Data generated or analyzed during this study can be are available from the corresponding author upon reasonable request.
\end{acknowledgments}
\bibliography{CJP8_SuperNova}%

\end{document}